\documentclass[useAMS,usenatbib]{mn2e}
\usepackage{graphicx}

\title[Gaussian Process Modelling of Asteroseismic Data]{Gaussian Process Modelling of Asteroseismic Data}
\author[B. J. Brewer and D. Stello]{B. J. Brewer$^{1,2}$\thanks{E-mail:
brewer@physics.usyd.edu.au} and D. Stello$^{1}$\\
$^{1}$Sydney Institute for Astronomy, School of Physics, The University of Sydney, 2006, NSW, Australia\\
$^{2}$School of Mathematics and Statistics, The University of New South Wales, 2052, NSW, Australia}
\begin{document}

\date{\today}
\pagerange{\pageref{firstpage}--\pageref{lastpage}} \pubyear{2002}
\maketitle

\label{firstpage}

\begin{abstract}
The measured properties of stellar oscillations can provide powerful constraints on the internal structure and composition of stars. To begin this process, oscillation frequencies must be extracted from the observational data, typically time series of the star's brightness or radial velocity. In this paper, a probabilistic model is introduced for inferring the frequencies and amplitudes of stellar oscillation modes from data, assuming that there is some periodic character to the oscillations, but that they may not be exactly sinusoidal. Effectively we fit damped oscillations to the time series, and hence the mode lifetime is also recovered. While this approach is computationally demanding for large time series ($>$ 1500 points), it should at least allow improved analysis of observations of solar-like oscillations in subgiant and red giant stars, as well as sparse observations of semiregular stars, where the number of points in the time series is often low. The method is demonstrated on simulated data and then applied to radial velocity measurements of the red giant star $\xi\textnormal{ Hydrae}$, yielding a mode lifetime between 0.41 and 2.65 days with 95\% posterior probability. The large frequency separation between modes is ambiguous, however we argue that the most plausible value is 6.3 $\mu$Hz, based on the radial velocity data and the star's position in the HR diagram.
\end{abstract}


\section{Introduction}
The study of stellar oscillations provides a powerful probe of the physical properties of stars. In particular, knowledge of the frequencies of many eigenmodes of a star can significantly constrain its internal structure and composition \citep[e.g.][]{2007CoAst.150..122H}. In practice, these frequencies are inferred from time series data of the star's radial velocity or intensity, which is analysed in order to determine the frequencies of the oscillation modes that contributed to the signal \citep{2007CoAst.150..106B}. The amount and quality of data has increased spectacularly over the past few years, mostly due to advances in instrumentation that were primarily intended for extrasolar planet searches \citep{1992PASP..104..270M}. Despite these advances, oscillation data on stars other than the Sun is still much more sparse and noisy, for obvious reasons. Most of this data is analysed with Fourier power spectrum methods inherited from helioseismology \citep[e.g.][]{1992A&A...257..287T, 2008MNRAS.389.1780J}. Recently, there has been growing interest in examining the fundamentals of data analysis techniques, and attempts to improve on these classical techniques have yielded modest, but non-negligible improvements to our ability to make use of time series data \citep[e.g.][]{2006MNRAS.371..935F, 2007ApJ...654..551B, 2007A&A...469..233R, stahn, 2008arXiv0811.3345G}. Separately, there has been a steady growth in interest in Bayesian Inference \citep{sivia} as the most consistent and natural way to model uncertainties. Thus, the approach described in this paper is Bayesian.

The idea behind Bayesian methods is to describe our knowledge by a probability distribution over the space of possible solutions we are considering. This probability distribution then gets updated to take into account the information contained in the data, in this case the time series data $\{y_i\}_{i=1}^N$. In asteroseismology, the possible solutions we consider are all possible values for the parameters of interest: the frequencies $\{\nu\}$ of the oscillation modes, their amplitudes $\{A\}$ and phases $\{\phi\}$, and the total number of modes, $m$. For brevity, we drop the braces hereafter; $A$, $\nu$ and $\phi$ now stand for arrays of amplitudes, frequencies and phases respectively. Any additional parameters (mode lifetime, for example) are denoted collectively by $\theta$. Before taking into account the data, we assign a {\it prior distribution} $p(\theta, m, \nu, A, \phi)$. We also probabilistically model the predictions for what data $y$ we expect to observe as a function of the parameters: $p(y|\theta, m,\nu,A,\phi)$, called the {\it sampling distribution}. Sometimes it is possible to marginalise out the amplitude and phase parameters, by integrating over them. This helps the computational search, because the algorithms only need to find good values for the frequencies, not the amplitudes and phases. This was done in a previous paper \citep[][hereafter B07]{2007ApJ...654..551B} by replacing the amplitudes and phases with sine-amplitudes and cosine-amplitudes; however, it is not possible for the model we introduce in this paper. Throughout this paper, we will be fitting both the frequencies and the amplitudes. It turns out (Section~\ref{simcov}) that we do not need phases in our model, so we will drop the phases hereafter.

Once the prior distribution and the sampling distribution have been specified, we have defined a joint probability distribution for the parameters and the data:
\begin{eqnarray}
p(\theta, m, \nu, A, y) = p(\theta, m, \nu, A)p(y|\theta, m, \nu, A)
\end{eqnarray}
This probability distribution describes what we know about the parameters and the data before we observe the actual data.
Once we learn the actual values of the data $y_{\textnormal{obs}}$, we update our probability distribution by deleting all hypotheses in the $(\theta, m, \nu, A, y)$ space that are now known to be false; i.e. we restrict our attention to the slice $y=y_{\textnormal{obs}}$. This gives the {\it posterior distribution} for the parameters given the data, describing our knowledge after updating to include the effect of the data. This is expressed by Bayes's theorem:
\begin{eqnarray}
p(\theta, m, \nu, A| y=y_{\textnormal{\small{obs}}}) \propto p(\theta, m, \nu, A)p(y|\theta, m, \nu, A)|_{y_{\textnormal{\small{obs}}}}\label{bayes}
\end{eqnarray}
The second factor in Equation~\ref{bayes} is the sampling distribution (probability distribution for the data as a function of the parameters) with the data fixed at the observed values; thus it is a function of the parameters only. Once the data have been fixed, it is commonly referred to as the {\it likelihood function}.

Many existing methods, including the Bayesian Analysis of B07, rely on the assumption that the observed time series is composed of a sum of sinusoidal signals, plus Gaussian noise - basically, this was our choice for the sampling distribution $p(y|\theta, m, \nu, A)$. Interestingly, the periodogram\footnote{Throughout this paper, the terms ``periodogram'' and ``power spectrum'' will be used interchangeably. It is also common to plot the square root of the periodogram, which is sometimes called the ``amplitude spectrum''.} can be proven to be a sufficient statistic from this same assumption, plus the assumption that the time series has complete phase coverage. What this means is that if we intend to infer the frequencies of the modes, no information is lost by reducing the time series to the periodogram \citep{bretthorst,2001AIPC..568..557G}, as long as the signal is purely sinusoidal and the time series has no significant gaps. Successful methods have been developed to infer frequencies and mode lifetimes by fitting to the power spectrum \citep[e.g.][]{2008A&A...488..705A}. However, the presence of stochastically excited modes and gaps in the data both break the assumptions required for the periodogram to be a sufficient statistic. Hence, {\it ideally}, when either of these conditions do not hold, we should work with the raw time series data.

Of course, it may be the case that using the power spectrum discards an insignificant amount of information, and the gain in convenience  of the power spectrum far outweights such theoretical concerns. This is certainly the case with solar data, and stellar data with good coverage and a long mode lifetime. It may be more generally true; however, this question is beyond the scope of this paper. In this paper we describe a new method to analyse time series data, taking into account the fact that the predicted signal from an oscillation mode is quasi-sinusoidal, and that the data may contain gaps, removing any concerns about information loss due to pre-processing via taking the power spectrum. This is done by making the choice for $p(y|\theta, m, \nu, A)$ as realistic as possible.

\section{Solar-Like Oscillations}
In the following, we will consider a single stochastically excited mode, and obtain a model for the sampling distribution $p(y|\theta, m, \nu, A)$ when $m=1$. Subsequently, we will expand this to include multiple oscillation modes and observational errors, in Section~\ref{independence}.
Solar-like oscillations are oscillations in main sequence or red giant stars that are continually damped and re-excited by turbulent convection, and therefore do not produce purely sinusoidal signals. 
For solar-like oscillations, the signal due to a single mode is modelled as a {\it damped and stochastically excited} oscillator with a driving force $f(t)$:
\begin{equation}
\frac{d^2y}{dt^2} + (2\pi\nu)^2 y + \frac{2}{\tau} \frac{dy}{dt} = \beta f(t)\label{ode}
\end{equation}
where $\tau$ is a damping timescale (the factor of two is introduced such that solutions to Equation~\ref{ode} without the driving force decay with an e-folding time of $\tau$), and $\beta$ is an amplitude constant for the driving force $f(t)$. If $f(t)$ is specified and the initial conditions $y(0)$ and $\dot{y}(0)$ are known, then Equation~\ref{ode} has a unique solution. However, as the term {\it stochastic excitation} suggests, $f(t)$ is only specified probabilistically. Throughout this paper, we will assume that $f(t)$ is unit variance white noise, so $f(t)$ at any time $t$ comes from a standard Gaussian distribution with mean 0 and standard deviation 1. $f(t_1)$ and $f(t_2)$ are independent for all distinct times $t_1 \neq t_2$. The white noise probability distribution that we assigned to $f(t)$ is an example of a {\it Gaussian Process} distribution. In general, a Gaussian Process is a probability distribution over a space of possible functions \citep{rasmussen, 2003itil.book.....M}; however a more general Gaussian Process may differ from white noise because the function value at different times may be correlated.

Since Equation~\ref{ode} is a linear ordinary differential equation, and $f(t)$ is assigned a Gaussian process distribution, we have implicitly also assigned a Gaussian process distribution for the value of the oscillating signal $y(t)$ at all times $t$. Whereas $f(t_1)$ and $f(t_2)$ are independent (for $t_1 \neq t_2$), the value of the oscillation signal at any two times, $y(t_1)$ and $y(t_2)$, are correlated, with the {\it covariance function} defined as
\begin{eqnarray}
C(t_i,t_j) = \left<y(t_i)y(t_j)\right> - \left<y(t_i)\right>\left<y(t_j)\right> \nonumber \\
= \left<y(t_i)y(t_j)\right>\label{cov}
\end{eqnarray}
where the expectatation value (mean) of $y$, $\left<y(t)\right>$, has been set to zero for all time. A typical signal obtained by solving Equation~\ref{ode} is shown in Figure~\ref{sim}. Clearly, the signal value at any given time is strongly correlated with the signal value at a time one period later, and anticorrelated with the value half a period later. However, the correlation is weaker than would be the case if the signal was purely sinusoidal. The entire signal displayed in Figure~\ref{sim} can be regarded as a single point sampled from a Gaussian Process distribution with mean function zero and a particular covariance function. Equation~\ref{ode} is capable of producing solutions with any initial phase, which is why we do not need phases in our model.
 
A useful property of Gaussian Processes is that the joint distribution for the function evaluated at a finite set of points (i.e. the signal $y(t)$ evaluated at the observation times) is a multivariate Gaussian, with covariance matrix given by the covariance function evaluated at the relevant points. Hence, the joint distribution for the value of the oscillation signal at a discrete set of $N$ times $\{t_1,t_2,...,t_N\}$ is:
\begin{equation}
p\left(y|A,\nu,\tau\right) = \frac{1}{\sqrt{\left(2 \pi\right)^N\det \mathbf{C}}} \exp\left(-\frac{1}{2}\mathbf{y}^T\mathbf{C}^{-1}\mathbf{y}\right)\label{gaussian}
\end{equation}
where $\mathbf{y} = \{y(t_1),y(t_2),...,y(t_N)\}$, the expectation values (means) of all of the $y$'s are zero, and $\mathbf{C}$ is a covariance matrix that implicitly depends on $A$, $\nu$ and $\tau$. Equation~\ref{gaussian} is the first step in the construction of a realistic $p(y|\theta,m, \nu, A)$: it would suit perfectly if we had noise-free data containing one mode, and if we knew how $\mathbf{C}$ depended on $A$, $\nu$ and $\tau$. The dependence of $\mathbf{C}$ on $A$, $\nu$ and $\tau$ is addressed in Section~\ref{simcov}, while the generalisation to noisy data with multiple modes occurs in Section~\ref{independence}.

Throughout this paper, no results depend on the choice of the origin for $t$, only relative times matter. In this case, the covariance function is said to be {\it stationary}. This implies that the covariance of $y(t_1)$ and $y(t_2)$ depends only on the difference between $t_1$ and $t_2$, and not on their absolute values:
\begin{equation}
C\left(t_1, t_2\right) = C(t_1,t_2) = C(|t_2 - t_1|) = C(\Delta t)
\end{equation}
where the symbol $C$ has been used to denote the covariance function, whether it takes one argument or two. This result is used to evaluate the elements of the covariance matrix $\mathbf{C}$.

\subsection{Use of Simulations to Determine the Covariance Function}\label{simcov}
In this section, we aim to find exactly how the covariance function for the signal depends on $A$, $\nu$ and $\tau$, for a single mode. The dependence on $A$ is trivial: if $y(t)$ comes from a Gaussian Process distribution with mean function zero and covariance function $C(\Delta t)$, then $A \times y(t)$ comes from a Gaussian Process distribution with mean zero and covariance function $A^2 C(\Delta t)$. Note that these quasi-sinusoidal signals do not have a strict amplitude like sinusoidal signals do, the amplitude $A$ is really just the expected standard deviation of the oscillation signal.

\begin{figure*}
\includegraphics[scale = 1.2]{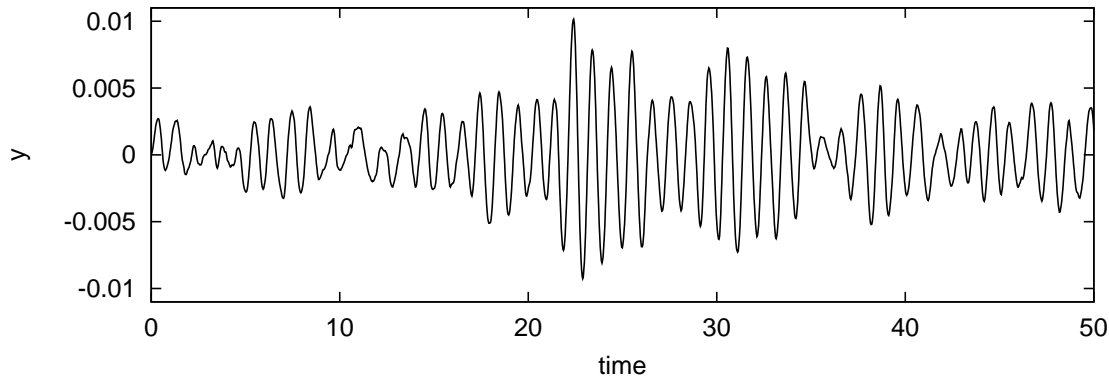}
\caption{Simulated signal from a single damped, stochastically excited oscillation mode: a solution to Equation~\ref{ode}. The frequency is 1 time unit, and the damping timescale (mode lifetime) $\tau$ is 10 time units. \label{sim}}
\end{figure*}

To investigate how the covariance function of a stochastically excited oscillation signal depends on frequency and mode lifetime, Equation~\ref{ode} was solved numerically using a fourth order Runge-Kutta algorithm with a timestep much smaller than the natural period of the oscillations. From a very long simulation, the covariance function for solutions of Equation~\ref{ode} was estimated by taking random pairs of times $t_1$ and $t_2$, and plotting the average value of $y(t_1)y(t_2)$ as a function of $\Delta t = |t_2 - t_1|$. A short section of the simulated time series is plotted in Figure~\ref{sim}, which clearly shows that while there is some periodic nature to the oscillations, the varying amplitude and phase changes would frustrate any simple modelling of the signal as sinusoidal waves - too many frequencies will be required in order to fit the data (B07).

\begin{figure*}
\includegraphics[scale = 0.75]{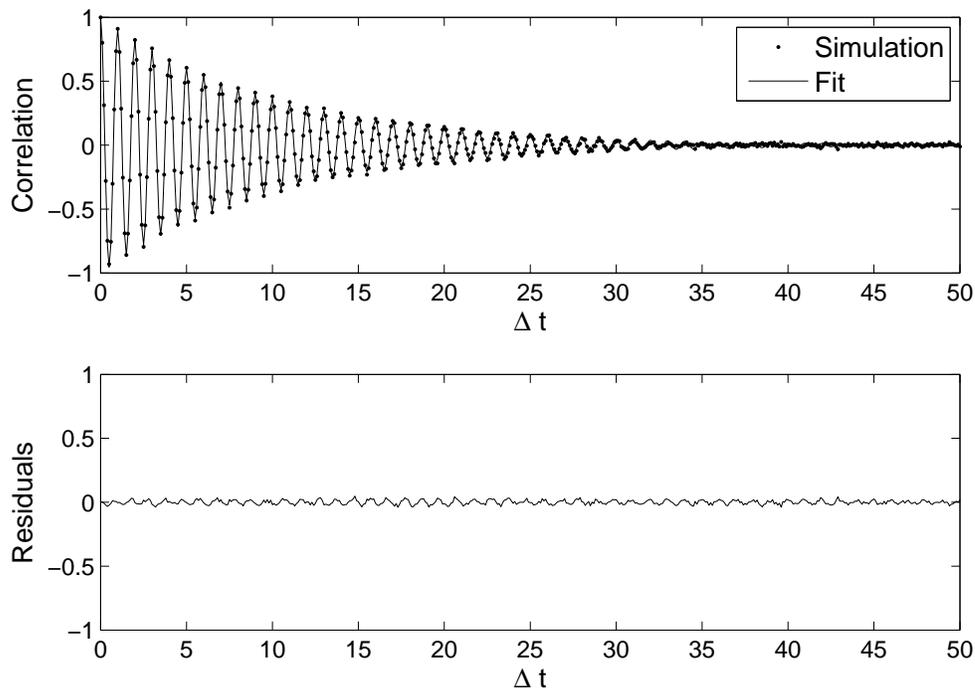}
\caption{The correlation function for solutions to the damped, stochastically driven oscillator problem, estimated from numerical simulations. The correlation is the same as the covariance function but with the variance scaled out. The top panel shows the result from simulations with an input frequency of 1 time unit and damping timescale (mode lifetime) of 10 time units. When parameterising the covariance function by $C(\Delta t) = A^2 \times \exp \left(-|\Delta t|/\tau'\right)\cos\left(2\pi\nu \Delta t\right)$, the best fit value of $\tau'$ is empirically found to be very close to the value of $\tau$ that was used to produce the data, so $\tau' \approx \tau$. The bottom panel shows the residuals after subtracting the fitted exponentially decaying cosine curve.\label{covfn}}
\end{figure*}

The estimated covariance function of the signal due to a single mode is shown in Figure~\ref{covfn}. It can be accurately modelled by a cosine curve multiplied by an exponential decay:
\begin{equation}
C(\Delta t) = A^2 \times \exp \left(-|\Delta t|/\tau'\right)\cos\left(2\pi\nu \Delta t\right)\label{cosdecay}
\end{equation}
where the decay timescale in the covariance function is empirically found to agree with $\tau$ to within 5 per cent (Figure~\ref{covfn}), and in practice we will take it as being equal. In the absence of stochastic excitation and damping, the covariance function would be just a cosine function, with frequency equal to the oscillation frequency. Hence, the effect of stochastic excitations and damping can be parameterised by the single parameter $\tau$ and its effect is to put an exponential decay factor into the covariance function for the signal. In the next subsection, this result is generalised to include multiple modes and observational errors.

\subsection{Addition of Independent Gaussian Processes}\label{independence}
Suppose there are two functions of time (for instance, the signal from two modes), $x(t)$ and $y(t)$ and our knowledge of these functions is described by independent stationary Gaussian Processes for each: with mean zero and covariance functions $C_x(\Delta t)$ and $C_y(\Delta t)$ respectively. If we are interested in the sum
\begin{equation}
z(t) = x(t) + y(t)
\end{equation}
then the sum is also a Gaussian process with mean zero and covariance
\begin{eqnarray}
C_z(\Delta t) = \left<z(t)z(t + \Delta t)\right>\nonumber\\ = \left<\left(x(t)+y(t)\right)\left(x(t+\Delta t)+y(t + \Delta t)\right)\right> \nonumber\\
= \left<x(t)x(t+\Delta t)\right> + \left<y(t)y(t+\Delta t)\right>\nonumber \\ + \left<y(t)x(t+\Delta t)\right> + \left<x(t)y(t+\Delta t)\right>
\end{eqnarray}
Since $x$ and $y$ are independent, the expectations of the last two terms are zero. Hence
\begin{eqnarray}
C_z(\Delta t) = \left<x(t)x(t+\Delta t)\right> + \left<y(t)y(t+\Delta t)\right> \nonumber \\ = C_x(\Delta t) + C_y(\Delta t)
\end{eqnarray}
Therefore, the covariance function for the sum of two independent Gaussian Processes is the sum of their individual covariance functions. This result can easily be extended to any number of Gaussian processes. Thus, if we are testing the hypothesis that there are many modes with various frequencies and amplitudes, and that there is also Gaussian noise in the data, the relevant covariance matrix (Equations~\ref{cov} and~\ref{gaussian}) is the sum of the covariance matrix for each mode (obtained from Equation~\ref{cosdecay}) and a diagonal covariance matrix for the noise, with the given measurement uncertainties used as the values for the noise standard deviation. In addition to these components, we included an ``extra noise'' signal to account for unmodelled errors, misquoted error bars, or correlated noise due to stellar effects that are not the oscillations of interest. The extra noise signal has an unknown standard deviation parameter $\sigma_{\textnormal{extra}}$ and a correlation timescale $\tau_\sigma$ for its exponentially decaying covariance function. Thus, $\sigma_{\textnormal{extra}}$ and $\tau_\sigma$ are additional parameters to be estimated from the data.

\section{Bayesian Inference}
We measure the signal at a discrete set of times $\{t_1,t_2,...,t_n\}$ with additive Gaussian noise of standard deviation $\{\sqrt{\sigma_1^2 + \sigma_{\textnormal{extra}}^2}, \sqrt{\sigma_2^2 + \sigma_{\textnormal{extra}}^2}, ..., \sqrt{\sigma_n^2 + \sigma_{\textnormal{extra}}^2}\}$, where $\sigma_1, \sigma_2 ...$ are the reported error bars on the  observations. The probability distribution for the total data set given all of the parameters (number of modes, their frequencies and amplitudes, the extra noise and its timescale) is Gaussian with covariance matrix formed by the sum of the covariance functions for each component (Section~\ref{independence}). Thus, we have now constructed the sampling distribution - the probability distribution for the observed data given the parameters of interest. The sampling distribution is the same as Equation~\ref{gaussian}, but where the covariance matrix $\mathbf{C}$ is the sum of the covariance matrices for each mode, the diagonal covariance matrix of the measurement errors, and the covariance matrix for the extra noise term. The posterior probability distribution for the parameters of interest given the data is then given by Bayes's theorem (Equation~\ref{bayes}), where $\theta = \{\tau, \sigma_{\textnormal{extra}}, \tau_\sigma\}$: i.e. the mode lifetime, extra noise standard deviation, and the correlation timescale for the extra noise term.

The previous sections described the sampling distribution, and thus the likelihood function, the second term in Equation~\ref{bayes}, in terms of Gaussian Processes. Now we must assign prior distributions for all of the parameters, i.e. the first term in Equation~\ref{bayes}. For simplicity, we chose the priors for all of the parameters to be independent of each other. In principle, this could be improved; for example, the expected amplitude of a mode is not the same at all frequencies.

The prior for the number of modes, $m$, was a uniform probability distribution ranging from 1 to a user-specified maximum number, which we took to be 200. The prior for the frequencies $\nu$ was a uniform distribution between a user-specified lower and upper limit: for the data sets discussed in this paper, these limits were 0 and 200 $\mu$Hz. The prior for the amplitudes $A$ was chosen to be an exponential distribution with unknown mean $\mu$, which effectively becomes yet another parameter to be inferred from the data. The priors for $\mu$, and the remaining parameters $\tau$, $\sigma_{\textnormal{extra}}$ and $\tau_\sigma$, all positive parameters, were chosen to be scale-invariant priors of the form $p(x) \propto 1/x$ between generous upper and lower limits. These priors correspond to uniform priors for the logarithm of the quantities, and is appropriate for positive parameters with unknown order of magnitude. Since the specification of the priors introduced an extra parameter $\mu$ to be inferred, the additional parameter vector is now extended to include $\mu$. Thus, $\theta = \{\tau, \sigma_{\textnormal{extra}}, \tau_\sigma, \mu\}$.

\section{Markov Chain Monte Carlo}
The posterior distribution can be effectively sampled using Markov Chain Monte Carlo (MCMC). In our implementation, we used the Metropolis algorithm \citep{neal}. Starting from a model with a single mode of arbitrary frequency and amplitude, and typical values for the additional parameters $\theta = \{\tau, \sigma_{\textnormal{extra}}, \tau_\sigma, \mu\}$, we propose to either add a mode (with its frequency and amplitude chosen from the prior), remove a mode, move a mode's frequency or amplitude, or shift the value of one of the additional parameters such as mode lifetime $\tau$. Then, the proposed change is accepted with a probability that depends on the relative likelihoods and prior probabilities of the current and the proposed model. Steps to models with higher posterior probability are always accepted, steps to models with a lower posterior probability are accepted with a probability given by the ratio of the posterior probability of the proposed model to the probability of the current one. If a proposed change to the model is rejected, the next model in the sequence is the same as the previous one. When this algorithm runs, the output of the code is a random sequence of models (sets of frequencies and amplitudes), each possibly slightly different from the last, where the diversity amongst the models is indicative of the uncertainty of any inference. To save memory, a subset of effectively independent models from this sequence may be used for any subsequent calculations. This is called ``thinning the chain''. For an introduction to MCMC see \citet{neal}, for a description within a context similar to this one, see B07. Unfortunately, the presence of the matrix inverse and determinant in the likelihood function (Equation~\ref{gaussian}) limits this algorithm to time series with less than $\sim$ 1500 points: even if the Cholesky decomposition is used to calculate det($\mathbf{C}$) and $\mathbf{C}^{-1}\mathbf{y}$ this still involves a calculation that takes time proportional to $N^3$, where $N$ is the number of points in the time series. For longer time series, other approaches are necessary; alternatively, approximations to Equation~\ref{gaussian} may be possible, but are beyond the scope of this paper. Additional efficiency can be obtained by using slice sampling \citep{slice} rather than Metropolis for the moving of frequencies and amplitudes.

\section{Simulated Data}
In this section, we demonstrate the use of our model on simulated data. To illustrate the method we show its output alongside output from other methods. We start with a simple case of a time series containing one mode and subsequently expand to several modes. We generated a long time series by numerically solving the ordinary differential equation~\ref{ode} for a single mode of frequency 100 $\mu$Hz and damping timescale (mode lifetime) 10$^5$ s = 1.1574 days or 10 oscillation periods. The amplitude of the signal was scaled to a standard deviation of 2 ms$^{-1}$ and then evaluated at 433 points in time, simulating 8 hours of nightly observations over an observing period of a month. In fact, the time stamps were the same as those from the $\xi\textnormal{ Hydrae}$ data observed by \citet{2002A&A...394L...5F}). Thus, the simulated data has the same window function as the actual $\xi\textnormal{ Hydrae}$ data. Measurement error was simulated by adding noise from a Gaussian distribution with a standard deviation of 2.5 m s$^{-1}$.

\begin{figure}
\includegraphics[scale=0.75]{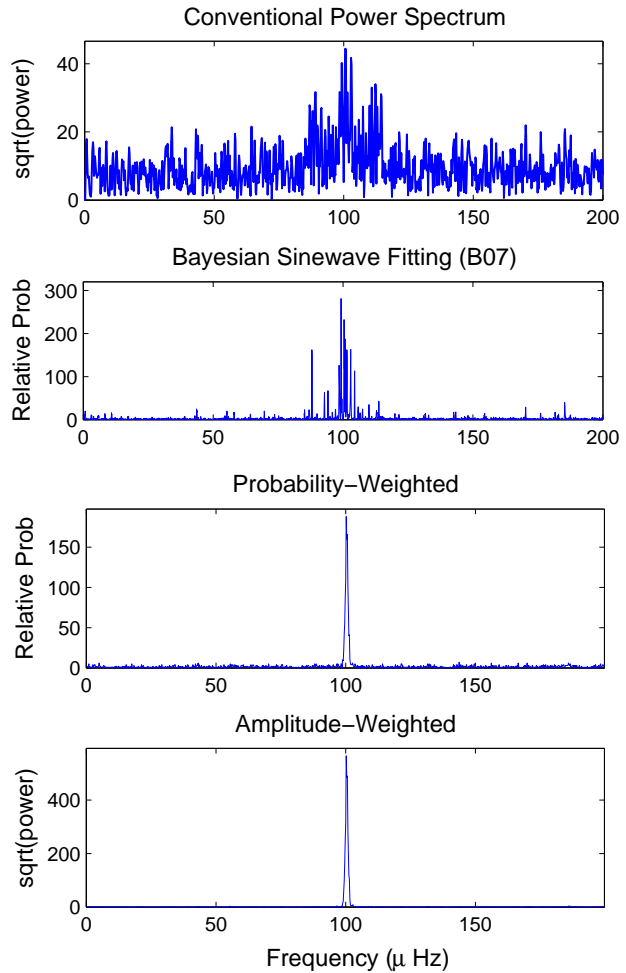}\caption{Results from analysing the single mode simulated data with various methods. Bayesian sine-wave fitting (2nd panel) can only explain the time series data by introducing many peaks, whereas the actual simulation contained only a single input frequency. Using the more realistic Gaussian Process likelihood, we find that a single frequency can explain the data (lower two panels).\label{onemode1}}
\end{figure}

The results obtained from analysing this data are displayed in Figure~\ref{onemode1}. The top panel shows the standard periodogram. In the 2nd panel, the results from Bayesian sine-wave fitting (B07) are shown. The B07 method is closely related to the iterative sine-wave fitting algorithm CLEAN, except that all frequencies are fitted simultaneously, and quantitative uncertainties are easily obtained. The MCMC approach of the B07 method actually returns a {\it sample} of fitted models, not a single one, however the results can be conveniently summarised by accumulating all detected frequencies into a single container, and then plotting a histogram of the frequencies. This histogram is what appears in the 2nd panel of Figure~\ref{onemode1}.  In the 3rd panel, a similar histogram is plotted of the output from running MCMC with the Gaussian Process likelihood introduced in this paper. Finally, the ``amplitude-weighted'' lower panel is a similar histogram, however in this case, before binning, each detected frequency is given a weight proportional to its amplitude. Thus, the 3rd panel indicates our confidence in the existence of a peak, while the 4th panel illustrates the estimated amplitude of each peak, and can be considered our version of an amplitude spectrum.

The MCMC run with the Gaussian Process likelihood took about one hour to complete (although an MCMC run never really finishes, just becomes more and more useful the longer it runs) on a modern PC with a 2 GHz dual core processor, compared to 15 minutes for the Bayesian sine-wave fitting and seconds for computing the periodogram. Note that the periodogram would need further processing, such as Lorentzian profile fitting \citep{2008arXiv0811.3345G}, to obtain a posterior distribution for the frequency of the mode. Doing this would result in a posterior similar to the 3rd panel of Figure~\ref{onemode1}, but with a slightly larger uncertainty due to the fact that the periodogram is not a sufficient statistic.

The approach outlined in this paper clearly identifies the presence of a mode with frequency 100.38 $\pm$ 0.48 $\mu$Hz. This is possible because the Gaussian process model takes into account at the outset the fact that the predicted signal due to a mode is not a pure sinusoid. Lacking this information, the sine-wave fitting approach is forced to introduce many peaks in order to explain the data (Figure~\ref{onemode1}). Note that the alias peaks at 90 and 110 $\mu$Hz are automatically removed by the Gaussian process model. They are only partially removed by the sinewave fitting, but would have been completely removed had the signal been truly sinusoidal.

\begin{figure}
\includegraphics[scale = 0.65]{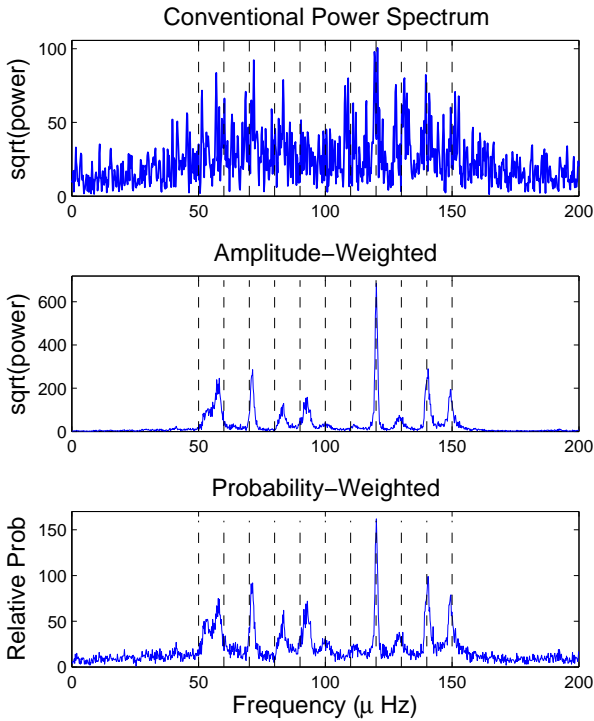}
\caption{Results from analysing simulated data with input frequencies from 50-150 $\mu$Hz in steps of 10 $\mu$Hz. Most of the modes are successfully identified, although some are spuriously shifted. The dashed lines indicate the true input frequencies.\label{multifreq_spectrum}}
\end{figure}

A further test of this method was done by testing it on simulated data containing many modes. Specifically, we generated simulated data from a star with 11 modes with frequencies ranging from 50-150 $\mu$Hz in steps of 10 $\mu$Hz. The time series contained 433 data points at the $\xi\textnormal{ Hydrae}$ times, as above. The results from analysing this simulated data set are shown in Figure~\ref{multifreq_spectrum}. While this result is less impressive than the single mode case, the algorithm has still successfully identified most of the input frequencies. There are some anomalies, such as the merging of the peaks at 50 and 60 $\mu$Hz, and the upward shift of the 80 $\mu$Hz mode. Note that the uncertainty about each frequency can be read off the width of the peaks in the bottom panel of Figure~\ref{multifreq_spectrum}. Whilst our method provides cleaner results than the raw periodogram, it is clearly not perfect. When interpreting results from this method, the summary plots like those shown in Figure~\ref{multifreq_spectrum} may be used as a guide, but the full output of the MCMC sample should be considered when the results are critical.

\begin{figure}
\includegraphics[scale = 0.6]{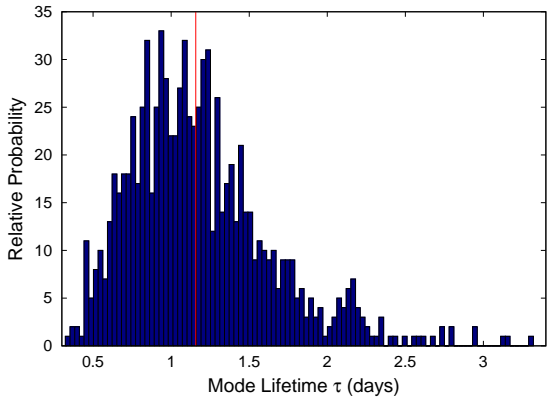}
\caption{The posterior distribution for mode lifetime $\tau$, given the simulated data set with 11 modes. The uncertainty is quite large, but comfortably contains the true input value $10^5 \textnormal{ s}=1.1574 \textnormal{ days}$.\label{modelifetime}}
\end{figure}

The mode lifetime $\tau$ can be measured using our analysis, as it is just another parameter that gets estimated by the MCMC. The posterior distribution for the mode lifetime is simply a histogram of the $\tau$ values encountered by the MCMC chain. For the multiple-mode simulated data, this distribution is shown in Figure~\ref{modelifetime}. The true input value of $10^5 \textnormal{ s}=1.1574 \textnormal{ days}$ is recovered, albeit with a large uncertainty, which is unsurprising given the time series is only 433 points in size. The distribution is asymmetric, largely due to our choice of a $1/\tau$ prior. Thus, conventional error bars are inappropriate. An alternative statement of uncertainty is the symmetric 95\% credible interval for the mode lifetime, which is [0.58, 2.10] days.

\section{The Large Separation and Mode Lifetime of $\xi\textnormal{ Hydrae}$}
We now turn to the real observations of $\xi\,$Hydrae. The first preliminary
analysis of the data presented by \citet{2002A&A...394L...5F} showed strong evidence
for solar-like oscillations based on the amplitude, the frequency
range, and the frequency separation of the extracted modes, which all
agreed with theoretical predictions.
They assumed that the mode lifetime was relatively long,
in accord with the theoretical calculations by
\citet{2002MNRAS.336L..65H} ($\tau\simeq17$ days), and hence their
analysis relied on the conventional power spectrum and iterative sine-wave fitting
(CLEAN). The value they found for the dominant frequency separation was
7.1 $\mu$Hz found between the strongest modes and 6.8 $\mu$Hz found from an
autocorrelation of the power spectrum in the region of excess power.
Subsequent studies of the same data including
extensive simulations by \citet{2004SoPh..220..207S} indicated that the
mode lifetime was significantly shorter than the
theoretical value ($\tau\simeq2$ days). This result was further confirmed
by \citet{2006A&A...448..709S} using an independent approach which also confirmed the
frequency separation found by \citet{2002A&A...394L...5F}, but they showed that the
precision by which the frequency separation could be established from the
data was low due to the short mode lifetime. In this
section we will apply our new Gaussian Process method to the
$\xi\,$Hydrae observations and compare our results with those found by the
previous studies.

\begin{figure}
\includegraphics[scale = 0.5]{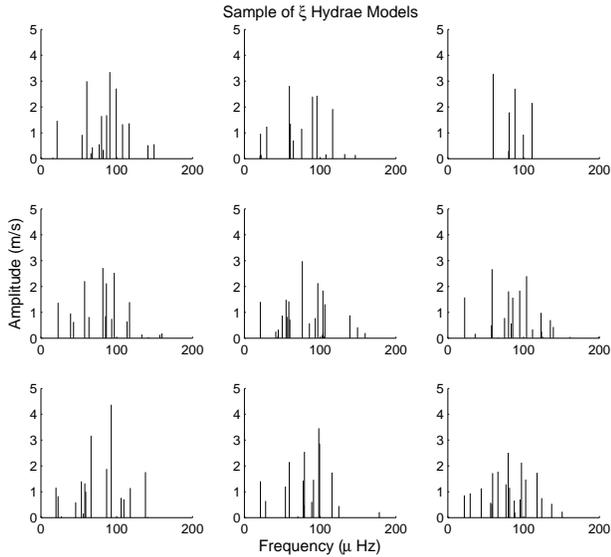}
\caption{A sample of models from the posterior distribution for the oscillation frequencies and amplitudes of $\xi\textnormal{ Hydrae}$. Clearly, the data and prior information do not uniquely determine the correct model. However, any question about the frequencies present can be answered probabilistically by calculating the fraction of the output models that have the property that is being tested for. The full sample is much larger than the nine models shown here.\label{xihya_sample}}
\end{figure}

\begin{figure}
\includegraphics[scale = 0.5]{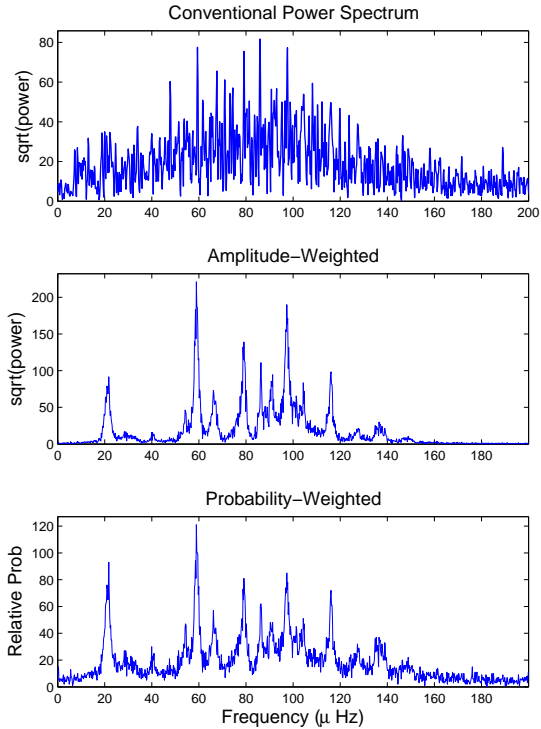}
\caption{Summarised results for the frequency spectrum of $\xi\textnormal{ Hydrae}$.\label{xihya_spectrum}}
\end{figure}

\begin{figure}
\includegraphics[scale = 0.6]{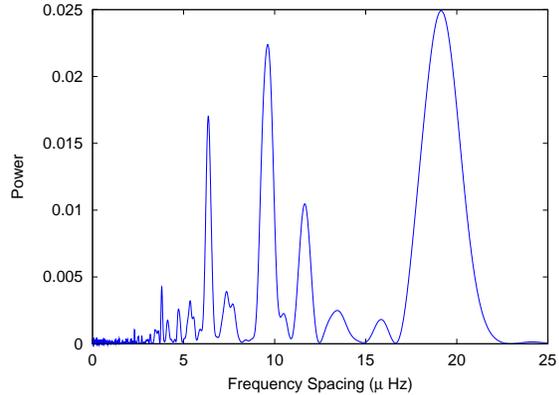}
\caption{Power spectrum of the estimated frequency spectrum (lower panel of Figure~\ref{xihya_spectrum}) of $\xi\textnormal{ Hydrae}$. A regular pattern to the frequencies of the modes should show up as a peak in this plot. There are high peaks at 6.3 $\mu$Hz, 9.6 $\mu$Hz and 19.1 $\mu$Hz.\label{sep}}
\end{figure}

The results from running our code on the $\xi\textnormal{ Hydrae}$ data are displayed in Figures~\ref{xihya_sample} and~\ref{xihya_spectrum}. The diversity of the models in Figure~\ref{xihya_sample} indicates that the uncertainties are quite large, and only a few modes are securely detected; this result agrees with the analysis of \citet{2006A&A...448..709S}. The large uncertainty about the frequencies is confirmed by the lower panel of Figure~\ref{xihya_spectrum} - the area under the curve over any frequency range is proportional to the probability that a mode exists within that range, yet even the peaks in this plot are only a factor of $\sim$ 2-3 higher than the background. There is some suggestion of a regular pattern to the peaks. To measure the large frequency separation, we took the power spectrum of the lower panel of Figure~\ref{xihya_spectrum} (a full Bayesian estimate of the large frequency separation, as done in B07, is prohibitive in this case) in order to search for periodicities. This power spectrum is displayed in Figure~\ref{sep} and shows at least three possible periodicities in the frequency pattern: one at 6.3 $\mu$Hz, another at 9.6 $\mu$Hz and a third peak at 19.2 $\mu$Hz (although this is simply a doubling of the 9.6 $\mu$Hz peak). Usually, if $l=0$ and $l=1$ modes are detected, the dominant separation of modes is half of the large frequency separation. This would imply that the large separation of $\xi\textnormal{ Hydrae}$ is 12.6, 19.2 or 38.4 $\mu$Hz.

However, from the classical stellar parameters (luminosity, mass, and effective temperature), the estimated large separation is 7.0 $\mu$Hz with about 10\% uncertainty, using the solar scaling presented in \citet{1995A&A...293...87K}. Also, a stellar pulsation model that goes through the star's position in the H-R diagram gives an average large spacing of 7.2 $\mu$Hz \citep{2002A&A...394L...5F}. Hence, the most plausible solution consistent with stellar astrophysics is that 6.3 $\mu$Hz is the large separation, not half of the large separation. Thus, we conclude that radial modes contributed most of the signal, non-radial modes may be excited with amplitudes below the detection threshold \citep{2006A&A...448..709S}. Although we have not formally modelled the uncertainty in the large separation, inspection of Figure~\ref{xihya_sample} shows that the uncertainty must be large. This uncertainty may be reduced with further observations, or perhaps by taking theoretical models of the star into account as prior information (B07). Our analysis used a uniform prior distribution for the frequencies, but if a large ensemble of plausible stellar models is produced, this would significantly reduce the range of possible frequency patterns and significantly improve the quality of the data analysis. Performing such an analysis is beyond the scope of this paper.

\begin{figure}
\includegraphics[scale = 0.6]{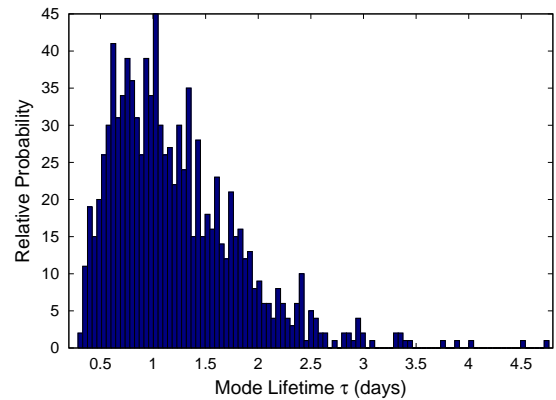}
\caption{The posterior distribution for mode lifetime $\tau$ for $\xi\textnormal{ Hydrae}$. The probability distribution for $\tau$ is asymmetric, but for $\log(\tau)$ it is approximately Gaussian. Hence, we estimate $\log_{10}(\tau/\textnormal{1 day}) = 0.03 \pm 0.21$ (1-sigma error bars). $\tau$ lies between 0.41 and 2.65 days with 95\% posterior probability.\label{modelifetime_xihya}}
\end{figure}

The posterior distribution for the mode lifetime of $\xi\textnormal{ Hydrae}$ is shown in Figure~\ref{modelifetime_xihya}. The mode lifetime is found to be very short, of order 1 day, albeit with a fairly large uncertainty. An estimate with 1-$\sigma$ error bars is $\log_{10}(\tau/\textnormal{1 day}) = 0.03 \pm 0.21$, and we find that $\tau$ lies between 0.41 and 2.65 days with 95\% posterior probability. This compares well with the result of \citet{2006A&A...448..709S} who estimated $\tau$ to be 2 days, but also with a large uncertainty. This mode lifetime remains much shorter than the theoretical predictions of 15-20 days \citep{2002MNRAS.336L..65H}.

\section{Conclusions}
In this paper, we have described a new Bayesian method for inferring the frequencies and amplitudes (with uncertainties on everything, including the number of modes present) of stellar oscillation modes from time series observations of the radial velocity or the intensity of the star. The method includes a Gaussian Process likelihood, which allows us to take into account the fact that the predicted signature of an oscillation mode is not exactly sinusoidal. Exactly how non-sinusoidal the oscillation signals are, is described by the mode lifetime, which is also estimated from the data, along with a measurement of the uncertainty in this value.

The method was implemented using a Markov Chain Monte Carlo algorithm and applied to two simulated data sets. As expected, the method removed the extra peaks caused by aliasing and the finite mode lifetime. We speculate that this method is, at the very least, comparable to the results obtained by fitting the power spectrum, but is more straightforward to interpret. Our method also avoids any concerns about information loss due to the fact that the power spectrum is not a sufficient statistic; whether this is of significant practical importance depends on the sampling of the time series, and the mode lifetime. For well-sampled time series and long mode lifetimes, information loss is not an issue.

Unfortunately, the method presented in this paper is computationally intensive due to the presence of a matrix inverse and determinant in the likelihood function. This limits its practical use to small time series with less than about 1500 points. For time series with 1500-15000 points, the Bayesian sine wave fitting approach (B07) is recommended, and for those with more than 15000 points, neither is computationally feasible; periodogram-based analysis is clearly the best choice here.

Applying the method to radial velocity data of the red giant $\xi\textnormal{ Hydrae}$, we found that the mode lifetime lies between 0.41 and 2.65 days with 95\% posterior probability. The large frequency separation was estimated to be either 6.3 $\mu$Hz or 9.6 $\mu$Hz, with the former being the most plausible given the star's position in the Hertzsprung-Russell diagram. C++ programs implementing the methods described in this paper (both the Gaussian Process and the sinewave fitting versions) are available upon request from B. J. Brewer.

\section*{Acknowledgments}
BJB would like to thank David MacKay for convincing me that a Gaussian Process approach would be feasible for this problem, and Matt Francis for bringing the Pictionary to my graduation party. The authors would like to thank Tim Bedding and Hans Bruntt for valuable discussion. BJB and DS acknowledge funding from the Australian Research Council. We also acknowledge the anonymous referee for their criticism, which helped us to improve the paper.

\label{lastpage}

\end{document}